\documentclass[10pt]{article}
\usepackage{times}
\usepackage{latexsym}
\usepackage{amsmath}
\usepackage{amssymb}
\usepackage{amsthm}

\def\mylabel#1{\label{#1}}
\theoremstyle{plain}
\newtheorem{satz}{Theorem}[section]
\newtheorem{prop}[satz]{Proposition}
\newtheorem{lem}[satz]{Lemma}
\newtheorem{cor}[satz]{Corollary}

\theoremstyle{definition}
\newtheorem{DEF}[satz]{Definition}
\newtheorem{bsp}[satz]{Example}
\newtheorem*{bsp*}{Example}

\newtheorem*{BEM*}{Remark}

\def\BNFdef{\;::=\;}
\def\BNFor{\mid}

\def\typeN{{\mathbf N}}
\def\typeBool{\mathbf{B}}
\def\typeD{\Diamond}
\def\typeTensor{\otimes}
\def\typeLinTo{\multimap}
\def\typeCross{\times}
\def\typeL#1{{\mathbf L}({#1})}
\def\typeT#1#2{\mathbf T({#1},{#2})}
\def\conPair#1#2{\left\langle{#1},{#2}\right\rangle}
\def\conItA#1{\left\{{#1}\right\}}
\def\conTit#1#2{\left\{{#1},{#2}\right\}}

\newcommand{\true}{\mathsf{t\kern-0.14em t}}
\newcommand{\false}{\mathsf{f\kern-0.14em f}}
\def\conTens#1#2{{\mathbf{\otimes}_{{#1},{#2}}}}
\def\nil#1{{\mathbf{nil}_{#1}}}
\def\cons#1{{\mathbf{cons}_{#1}}}
\def\emptyTree#1#2{\mathbf{leaf}_{{#1},{#2}}}
\def\consTree#1#2{\mathbf{tree}_{{#1},{#2}}}
\def\len#1{\left|{#1}\right|}
\def\pollen#1{\vartheta\left({#1}\right)}
\def\polleq{\preccurlyeq}
\def\suppol#1{\mathrm{sup}_{\polleq}\left\{{#1}\right\}}

\def\typed#1#2#3{{#1}\vdash{#2}^{#3}}
\def\Rule#1#2{\begin{array}{c}{#1}\\ \hline {#2}\\{}\\ \end{array}\qquad}

\def\max#1{{\mathrm{max}\left\{{#1}\right\}}}
\def\min#1{{\mathrm{min}\left\{{#1}\right\}}}
\def\card#1{\left|{#1}\right|}
\def\FV#1{\mathrm{FV}\left({#1}\right)}
\def\redel{\mapsto}
\def\red{\to}
\def\linvec#1{\vec{#1}}
\def\subst#1#2{[{#1}/{#2}]}
\def\natnumb{{\mathbb N}}
\def\N{\natnumb}        
 
\def\Np{\N^{\mathrm{poly}}}
\def\X{X}                 
\def\leqbsp#1{\lhd_{#1}} 
\def\typeI{\iota}
\def\Izero{\circ}
\def\Isuc#1{\mathfrak{s}_{#1}}
\def\IsucA{\Isuc{0}}
\def\IsucB{\Isuc{1}}
\def\Ipred{\mathbf{p}}
\def\Iiszero{\mathbf{iszero}}
\def\Ifirstdigit{\mathbf{head}}
\def\varTypeA{\tau}
\def\varTypeB{\rho}
\def\varTypeC{\sigma}
\def\varTerma{a}
\def\varTermt{t}
\def\varTerms{s}
\def\varTermr{r}
\def\varTermh{h}
\def\varTermd{d}
\def\funF{\mathfrak{f}}
\def\funG{\mathfrak{g}}

\def\varx{x}
\def\vary{y}
\def\varz{z}
\def\varl{l}
\def\varp{p}

\def\varCon{c}

\def\varL{\ell}
\def\varpolf{f}
\def\varpolg{g}
\def\varnatk{k}
\def\varnatn{n}
\def\varnati{i}
\def\varnatj{j}
\def\varnatm{m}

\def\varnatN{\frak N}
\def\varKon{\Gamma}

\def\bbewn#1{\begin{proof}[#1]}
\def\bbew{\begin{proof}}
\def\ebew{\end{proof}}
\def\qedsymbol{{\ \vbox{\hrule\hbox{%
   \vrule height1.3ex\hskip0.8ex\vrule}\hrule}}\par}

\def\Ind#1{Induction on $ #1 $}
\def\varnatN{N}
 
\newcommand{\inquotes}[1]{``{#1}''}
              
\begin{document}

\title{A syntactical analysis of non-size-increasing polynomial time
       computation}
\author{Klaus Aehlig\thanks{supported by the ``Graduiertenkolleg Logik
in der Informatik'' of the ``Deutsche Forschungsgemeinschaft''} 
 \and Helmut Schwichtenberg}
\date{~}
\maketitle

\begin{abstract}
A syntactical proof is given that all functions definable in a
certain affine linear typed $\lambda$-calculus with iteration in all
types are polynomial time computable.  The proof provides explicit
polynomial bounds that can easily be calculated.
\end{abstract}

\begin{section}{Introduction}
Recent research
\cite{BellantoniCook92,BellantoniNigglSchwichtenberg00,Hofmann98b} has
provided many characterizations of {\tt Ptime} (the class of all
functions computable in polynomial time) by means of appropriate
restrictions of the terms in G\"odel's~$T$~\cite{Goedel58}.
Consider the following definition of an exponentially growing function.
\begin{align*}
\begin{split}
\mathtt{double}([\;]) &:= [\;] 
\\
\mathtt{double}([\,a \mid \ell\,]) &:= [\,a, a \mid \mathtt{double}(\ell)\,]
\end{split}
\quad
\begin{split}
\mathtt{exp}([\;]) &:= [\true] 
\\
\mathtt{exp}([\,a \mid \ell\,]) &:= \mathtt{double}(\mathtt{exp}(\ell))
\end{split}
\end{align*}
Approaches based on predicative
recursion~\cite{Simmons88,BellantoniCook92,Leivant91} argue that 
the exponential growth in this example
is due to the way $\mathtt{double}$ is called: the previous value 
$\mathtt{exp}(\ell)$ of the outer
recursion is the recursive argument to the inner recursion.

Although such approaches can capture all
\emph{functions} computable in polynomial time, many natural
\emph{algorithms} are excluded, particularly if they involve nested 
recursions.  Standard examples are sorting algorithms like insertion
sort, which has a similar recursive structure as {\tt exp}:
\begin{alignat*}{2}
\mathtt{insert}(a,[\;]) &:= [a] \\
\mathtt{insert}(a,[\,b \mid \ell\,]) 
&:= [\,b^{\prime} \mid \mathtt{insert}(a^{\prime},\ell)\,]
&\quad&\text{$a^{\prime}, b^{\prime}$ a permutation of $a,b$
with $b^{\prime}\le a^{\prime}$}\\[1ex]
\mathtt{sort}([\;]) &:= [\;] \\
\mathtt{sort}([\,a \mid \ell\,]) &:= \mathtt{insert}(a,\mathtt{sort}(\ell))
\end{alignat*}
Caseiro~\cite{Caseiro97} studied many related
examples and reached some (partially semantic) criteria 
for algorithms in order to ensure polynomial time complexity.  For the
insertion sort algorithm the essential point is that
$\mathtt{insert}$ does not increase the size of its input.
Hofmann~\cite{Hofmann99} took up this line of research and
formulated a new term system with a special type $\typeD$ of tokens
that accommodates nested recursion, but only allows to 
define non-size-increasing {\tt Ptime} functions.
The basic idea is that if
a function increases the size of its argument (like a successor
function), then one has to ``pay'' for the increase by providing a
\inquotes{token}, i.e., a term of type $\typeD$.

Hofmann~\cite{Hofmann99} proved this result
by means of inherently semantic concepts,
such as the set-theoretic interpretation of terms.
We present a new proof of his main result, which apart from being
simpler provides more insight and yields an explicit construction of
time bound polynomials. 
Although Hofmann's proof is constructive and thus also contains
bounds, these are deeply hidden.

The method developed here has several benefits:
\begin{itemize}
\item  A reduction relation is defined in such a way 
that the term system is closed under it.
Therefore calculations can be
performed \emph{within} the system.
\item We not only show that every definable function is polytime, but
give explicit 
polynomial bounds for the number of reduction steps
that can be determined easily for any
given term.
\item Hofmann's semantical size measure~\cite[\S3.2]{Hofmann99}
(minimal upper bound of growth) is replaced by the syntactic concept of the
number of free variables. Hence the role of the
$\typeD$-type becomes more transparent, as we will show (in
Lemma~\ref{lem-term-dia}) that there are no closed terms of this type.
\end{itemize}

A preliminary version~\cite{AehligSchwichtenberg00} of this work has
already been published.  Apart from giving more technical
details and elaborated proofs, the following aspects are added:
\begin{itemize}
\item The previous estimate of the number of reduction steps referred
to a fixed reduction strategy.  Here we show that this requirement can
be relaxed somewhat, without loosing the sharp estimate on the length
of reduction sequences.  The main tool is an appropriate modification
of the size measure $\pollen{\cdot}$.
\item It is shown how the approach covers more complex data structures such
as binary labeled trees, for which iteration involves two recursive
calls.
\end{itemize}

Hofmann's paper \cite{Hofmann99} is of course the starting point of
ours.  In the journal version \cite{Hofmann00} some new aspects are
added, among others the definition and justification of a similar
system which captures \texttt{Pspace}, and an operator for
divide-and-conquer recursion.  We shall show in Section~\ref{SS:Trees}
that our simplified approach can deal with the latter; however, it
is not known whether it also suffices for Hofmann's characterization
of \texttt{Pspace}.

Jones \cite{Jones01}, in a paper called \inquotes{The expressive power
of higher-order types or, life without CONS}, directly relates
programming languages and complexity theory.  In a previous paper
\cite{Jones99} (see also \cite{Jones97}) he had characterized the
power of first-order read-only programs in complexity terms;
the results are extended~\cite{Jones01} to arbitrary data orders and
tail recursive programs.  As Jones notes, the root for his
\texttt{Ptime} result (for data order $0$) is an early paper of Cook
\cite{Cook71} on \inquotes{two-way auxiliary pushdown automata}.
However, no characterization of \texttt{Ptime} in terms of recursion
with unlimited data orders is obtained.

Jones work is certainly related to the present one: in his approach it
is important whether constructors of structured data are allowed or
not.  Moreover, Jones' paper is written from a broader programming
point of view, giving special emphasis to the program control
structure (general recursion, tail recursion or primitive recursion).
The present paper (for simplicity) concentrates on the effects of
higher order primitive recursion (i.e.\ G\"odel's $T$).

Leivant and Marion \cite{LeivantMarion93a} give a $\lambda$-calculus
characterization of \texttt{Ptime}.  The main novelty is the use of
the concept of \inquotes{tiers} in the sense of different
representations of data: as words (i.e.\ terms of base type), or as
Church-like abstraction terms (expressions of higher type).  This
leads to a nice characterization of \texttt{Ptime}.  Leivant
\cite{Leivant99} gives particularly elegant proofs of
characterizations of Jones \cite{Jones99} for \texttt{Ptime} and of
Beckmann and Weiermann \cite{BeckmannWeiermann00} for the elementary
functions.  He also treats a control-based sub-calculus of the system
in \cite{BellantoniNigglSchwichtenberg00}, and ---~more importantly~---
obtains a characterization of \texttt{Pspace} along the same lines.
This is achieved by the notions of an \inquotes{input driven} term,
i.e., a term with the property that no recursion argument has a free
variable bound in the term, and the more special notion of a
\inquotes{solitary} term, where every higher type abstracted variable
has at most one occurrence in the kernel.  The solitary terms are the
ones used for his characterization of \texttt{Ptime}. As with all
extensional characterizations of \texttt{Ptime}, the question remains
as to whether interesting \texttt{Ptime} algorithms can be represented
directly.
\end{section}

\begin{section}{Types and terms}
\begin{DEF}[Finite types]\label{def-lin-types}
The set of linear types is defined inductively as
\[
\varTypeB,\varTypeA\BNFdef \typeD\BNFor\typeBool
                            \BNFor\varTypeA\typeLinTo\varTypeB
                            \BNFor\varTypeA\typeTensor\varTypeB\BNFor
                            \varTypeA\typeCross\varTypeB\BNFor\typeL{
	                        \varTypeA}
\]
\end{DEF}
So types are built from the base type
 $\typeBool$ of booleans and a special type
$\typeD$. 
$\typeL{\varTypeA}$ denotes the list-type over $\varTypeA$.
The type of binary numerals, as used by
Hofmann~\cite{Hofmann99}, can be defined as $\typeL{\typeBool}$.
The type $\typeD$ may be interpreted as a pointer to free
memory~\cite{Hofmann01}; 
Lemma~\ref{lem-term-dia} will show that no closed terms of type
 $\typeD$ exist.
Canonical terms of data-types may still contain free variables
of that type.
The type
$\typeD\typeLinTo\varTypeA\typeLinTo\typeL{\varTypeA}
 \typeLinTo\typeL{\varTypeA}$
of the $\cons{\varTypeA}$ function
together with the linear typing discipline guarantees that the
length of lists and the number of free variables coincide.

$\varTypeA\typeTensor\varTypeB$ and $\varTypeA\typeCross\varTypeB$ 
both represent ordered pairs. However, the linear meaning of
those types is different. Each component of a tensor product
$\varTypeA\typeTensor\varTypeB$ can be used,
whereas in the case of an ordinary pair 
one can choose only one component to
be used.  As can be seen from Definition~\ref{outer-red}, $\typeTensor$
corresponds to $\typeLinTo$, while $\typeCross$ corresponds to
$\typeBool$.

Terms are built from variables, denoted by $\varx,\vary,\varz$, and
typed
constructor symbols $\varCon$.  Each variable has a type, and it is
assumed that there are infinitely many variables of each type.  The
notation $\varx^{\varTypeA}$ should express that the variable $\varx$
has type $\varTypeA$.

\begin{DEF}[Terms]
The set of terms is inductively defined by
\[
\varTermr,
\varTerms,\varTermt\BNFdef\varx^{\varTypeA} \BNFor \varCon \BNFor
\lambda\varx^{\varTypeA}.\,\varTermt \BNFor
\conPair{\varTermt}{\varTerms} \BNFor
\varTermt\varTerms \BNFor
\conItA{\varTermt}
\]
\end{DEF}

These terms should be seen as our ``raw syntax''; only correctly typed
terms (in the sense of Definition~\ref{typed-terms}) represent
meaningful functions.

The idea for the $\conItA{\cdot}$ term
construct is taken from Joachimski and Matthes~\cite{JoachimskiMatthes99}.  
Terms of this
form appear only as arguments in an $\typeL{\varTypeA}$
elimination.  This explicit marking of the step term in iteration
constructs is not only technically convenient but also allows to
directly read off the time complexity:
the degree of the bounding polynomial is the nesting depth of braces
and the symbols within $\varnatn$ braces contribute to the coefficient of
$\X^{\varnatn}$ (see Definition~\ref{def-pollen} for details).

The notations $\varTermt\linvec{\varTerms}$ and
$\lambda\linvec{\varx}\,^{\linvec{\varTypeA}}.\,\varTermt$ are defined
as usual, so $\lambda\varx^{\varTypeA},\vary^{\varTypeB}.\,\varTermt$
is an abbreviation for
$\lambda\varx^{\varTypeA}.(\lambda\vary^{\varTypeB}.\,\varTermt)$.
Terms that only differ in the naming of bound variables are
identified.
\begin{DEF}[Constructor symbols]\mylabel{def-kon-sym}
The constructor symbols and their types are
\[
\begin{array}{ll}
\true & \typeBool \\
\false\qquad & \typeBool \\
\nil{\varTypeA} & \typeL{\varTypeA} \\
\cons{\varTypeA} & \typeD\typeLinTo\varTypeA\typeLinTo\typeL{\varTypeA} 
					\typeLinTo\typeL{\varTypeA} \\
\conTens{\varTypeA}{\varTypeB} & \varTypeA\typeLinTo\varTypeB\typeLinTo
                                \varTypeA\typeTensor\varTypeB \\
\end{array}
\]
\end{DEF}

A \emph{context} is a set of variables.  For two contexts $\varKon_1$
and $\varKon_2$ the notation $\varKon_1,\varKon_2$ stands for the
union, expressing that $\varKon_1$ and $\varKon_2$ are disjoint.  
We also write $\varx^{\varTypeA}$ for the singleton
context $\{\varx^{\varTypeA}\}$.

The next definition states which terms of our raw syntax are correctly
typed.  $\typed{\varKon}{\varTermt}{\varTypeA}$ is to be read as
``$\varTermt$ is a typed term of type $\varTypeA$ with free variables
in $\varKon$''. 

\begin{DEF}[$\typed{\varKon}{\varTermt}{\varTypeA}$]
\mylabel{typed-terms}
The relation $\typed{\varKon}{\varTermt}{\varTypeA}$ is inductively defined
as follows:

$
\begin{array}{p{0.75\textwidth}l} $
\Rule{}{\typed{\varKon,\varx^{\varTypeA}}{\varx}{\varTypeA}}  
$& (Var) \end{array}  $ \par $   \begin{array}{p{0.75\textwidth}l} $
\Rule{\varCon\mbox{~of type~}\varTypeA}{\typed{\varKon}{\varCon}{\varTypeA}}  $&
(Const) \end{array}  $ \par $   \begin{array}{p{0.75\textwidth}l} $
\Rule{\typed{\varKon\cup\{\varx^{\varTypeA}\}}{\varTermt}{\varTypeB}}
     {\typed{\varKon}{(\lambda\varx^{\varTypeA}.\,t)}{\varTypeA\typeLinTo
                                                    \varTypeB}}  $&
(\typeLinTo^{+}) \end{array}  $ \par $   \begin{array}{p{0.75\textwidth}l} $
\Rule{\typed{\varKon_1}{\varTermt}{\varTypeA\typeLinTo\varTypeB}\qquad
      \typed{\varKon_2}{\varTerms}{\varTypeA}}
     {\typed{\varKon_1,\varKon_2}{(\varTermt\varTerms)}{\varTypeB}}  $&
(\typeLinTo^-) 
\end{array}   $ \par $   \begin{array}{p{0.75\textwidth}l} $
\Rule{\typed{\varKon}{\varTermt}{\varTypeA}\qquad
      \typed{\varKon}{\varTerms}{\varTypeB}}
     {\typed{\varKon}{\conPair{t}{s}}{\varTypeA\typeCross\varTypeB}}  $&
(\typeCross^+) 
\end{array}   $ \par $   \begin{array}{p{0.75\textwidth}l} $
\Rule{\typed{\varKon}{\varTermt}{\varTypeA\typeCross\varTypeB}}
     {\typed{\varKon}{(\varTermt\true)}{\varTypeA}}  $&
(\typeCross^-_1) \end{array}  $ \par $   \begin{array}{p{0.75\textwidth}l} $
\Rule{\typed{\varKon}{\varTermt}{\varTypeA\typeCross\varTypeB}}
     {\typed{\varKon}{(\varTermt\false)}{\varTypeB}}  $&
(\typeCross^-_0)
\end{array}  $ \par $   \begin{array}{p{0.75\textwidth}l} $
\Rule{\typed{\varKon_1}{\varTermt}{\typeBool}\qquad
      \typed{\varKon_2}{\varTerms}{\varTypeA}\qquad
      \typed{\varKon_2}{\varTermr}{\varTypeA}}
     {\typed{\varKon_1,\varKon_2}{(\varTermt\conPair{\varTerms}{\varTermr})}
            {\varTypeA}}  $&
(\typeBool^-) \end{array}  $ \par $   \begin{array}{p{0.75\textwidth}l} $
\Rule{\typed{\varKon_1}{\varTermt}{\varTypeA\typeTensor\varTypeB} \qquad
      \typed{\varKon_2,\varx^{\varTypeA},\vary^{\varTypeB}}{\varTerms}
            {\varTypeC}}
	 {\typed{\varKon_1,\varKon_2}{(\varTermt(\lambda\varx^{\varTypeA},\vary^{
            \varTypeB}.\varTerms))}{\varTypeC}}  $&
(\typeTensor^-) \end{array}  $ \par $   \begin{array}{p{0.75\textwidth}l} $
\Rule{\typed{\varKon}{\varTermt}{\typeL{\varTypeA}}\qquad
      \typed{\emptyset}{\varTerms}
            {\typeD\typeLinTo\varTypeA\typeLinTo\varTypeB\typeLinTo\varTypeB}}
     {\typed{\varKon}{(\varTermt\conItA{\varTerms})}{\varTypeB\typeLinTo
       \varTypeB}}  $&
(\typeL{\varTypeA}^-)
\end{array}$
\end{DEF}

It is crucial that the step term in 
$(\typeL{\varTypeA}^-)$ is closed. Otherwise it would not be possible
to define reduction rules in such a way that typed terms reduce to typed
terms (Lemma~\ref{lem-typ-red}),
since free variables in the step terms would be duplicated.

The typing system is based on elimination rules: For
every type different to $\typeD$ there is a rule describing how to
use terms of this type; in all these rules the elimination is written
as application%
\footnote{
It should be noted that the same syntax would appear in the standard
implementation in the untyped $\lambda$-calculus. For example one should
think of $\typeTensor$ as being $\lambda x,y,z. z x y$.
}.
For the right hand side of such an application
a notation is chosen that expresses the computational behavior
of the terms used for the elimination. This avoids duplication of syntax,
e.g., with the pair $\conPair{\varTermt}{\varTerms}$ we have a notation 
expressing
that exactly one of the terms $\varTermt$ and $\varTerms$ will be
needed. This syntax is also used for the 
``if\dots then\dots else\dots'' construct
$\varTermt\conPair{\varTerms}{\varTermr}$. 

Data-types, i.e., types from which all the stored
information can be retrieved
by a single use of an object of this type, are introduced by
constants. 
In the system we have two forms of abstraction with special
introduction rules: the $\lambda$-abstraction
and the pair. Note that the pair really is an abstraction, as only
part of the stored information can be retrieved by a single use. The
tensor $\varTypeA\typeTensor\varTypeB$
however is a data-type as a single use can access both
components, hence we have a constant $\conTens{\varTypeA}{\varTypeB}$ 
to introduce it.

This way of introduction allows the relatively simple
Definition~\ref{def-term-closure} of the reduction relation expressing
that we may reduce within data, but not under abstractions.

Immediately obvious from the definition is
\begin{prop}[Weakening]\mylabel{prop-term-weak}
If $\typed{\varKon}{\varTermt}{\varTypeA}$
and $\varKon\subset\varKon^{\prime}$ then
$\typed{\varKon^{\prime}}{\varTermt}{\varTypeA}$.
\end{prop}

Rule $(\typeLinTo^+)$ could as well have been written
$$
\Rule{\typed{\varKon,\varx^{\varTypeA}}{\varTermt}{\varTypeB}}
{\typed{\varKon}{(\lambda\varx^{\varTypeA}.\,t)}{\varTypeA\typeLinTo
\varTypeB}}
$$
requiring that the bound variable does not occur in the context
afterwards. In our formulation it is easier to recognize
weakening as a derived rule. As we identify $\alpha$-equal
terms, weakening holds in the alternative formulation as well and both
formulations are in fact equivalent.

It might seem odd that in this calculus a typed term can have an
untypable subterm $\conItA{\varTermt}$.  An obvious definition
would introduce a new form of application, e.g., if $\varTermr,
\varTerms$ are terms then so is $\varTermr \conItA{\varTerms}$,
rather then a new term former $\conItA{\varTerms}$.  However,
the present approach is technically more convenient, since it allows
the simple classification of terms according to their head 
form (cf.\ Lemma~\ref{lem-kopfform}).

It should be noted that with some technical overhead one could resolve
this problem by separating terms and so-called elimination terms,
as in work of Joachimski and Matthes~\cite{JoachimskiMatthes99}.
But in the present rather simple situation this seems to be an
overkill.

The notation $\varTermt^{\varTypeA}$ expresses that there is
a $\varKon$ such that $\typed{\varKon}{\varTermt}{\varTypeA}$.
The smallest such $\varKon$ is called the set of free variables.
By induction on the
definition of $\typed{\varKon}{\varTermt}{\varTypeA}$ one easily verifies

\begin{lem}\mylabel{lem-term-typ}
For each term $\varTermt$ there is at most one type $\varTypeA$ such that
$\varTermt^{\varTypeA}$. In this case there is a smallest
$\varKon$ with $\typed{\varKon}{\varTermt}{\varTypeA}$ which
coincides with
the set of free variables of $\varTermt$ in the
usual sense.
\end{lem}

As already mentioned, the main restriction of the typing calculus
is linearity, that
is, in an application $\varTermt_1\varTermt_2$ the free variables of
$\varTermt_1$ and $\varTermt_2$ have to be disjoint. This is stated
explicitly in the rule $(\typeLinTo^-)$, but holds for all other
``applications'' as well. More precisely 

\begin{lem}[Linearity]\mylabel{lem-inversion}
Assume $(\varTermt_1\varTermt_2)^{\varTypeA}$. Then either
$\varTermt_2=\conItA{\varTermt_2'}$ with closed $\varTermt_2'$
or else there are types $\varTypeA'$ and $\varTypeA''$ and disjoint
contexts $\varKon_1,\varKon_2$ such that
$\typed{\varKon_1}{\varTermt_1}{\varTypeA'}$
and $\typed{\varKon_2}{\varTermt_2}{\varTypeA''}$
\end{lem}
\begin{proof}[Proof%
\footnotemark]
\footnotetext{It should be noted that this is just an inspection of the
rules and in particular \emph{no} induction is needed.}
The last rule of a derivation
of $\typed{\varKon}{(\varTermt_1\varTermt_2)}{\varTypeA}$ must be one
of $(\typeLinTo^-)$, $(\typeCross_0^-)$, $(\typeCross_1^-)$,
$(\typeBool^-)$, $(\typeTensor^-)$ or $(\typeL{\varTypeA}^-)$. In each
of these cases the claim is trivial. For example, in the case
$(\typeTensor^-)$ we conclude by two applications of $(\typeLinTo^+)$
that $\typed{\varKon_2}{(\lambda\varx^{\varTypeA},\vary^{\varTypeB}.
\varTerms)}{\varTypeA\typeLinTo\varTypeB\typeLinTo\varTypeC}$.
\end{proof}

The fact that all eliminations are written as applications ensures
that all typed terms have a uniform appearance. As can easily be
verified by induction on the definition of the
$\typed{\cdot}{\cdot}{\cdot}$ relation, we have

\begin{lem}[Head form] \mylabel{lem-kopfform}
Assume $\typed{\varKon}{\varTermt}{\varTypeA}$.  Then $\varTermt$ is
of the form $\varx\linvec{\varTermt}$, $\varCon\linvec{\varTermt}$,
$(\lambda\varx^{\varTypeB}.\,\varTerms)\linvec{\varTermt}$ or
$\conPair{\varTerms}{\varTermr}\linvec{\varTermt}$.
\hfill \qedsymbol
\end{lem}

It should be noted that this lemma, although technically trivial,
turns out to be crucial for the further development: As we are taking
vector notation serious and only have constants for the
introduction of data-types, case distinction
according to this lemma (and further according to the constant
$\varCon$ in the case $\varCon\linvec{\varTermt}$) is essentially
case distinction according to the ``form of the next canonical
redex'', without the need of defining such a notion.
\end{section}

\begin{section}{Reductions}
The reduction rules to be defined are all correct with
respect to the set-theoretic semantics (cf.\ 
Hofmann~\cite[\S2.1]{Hofmann00}).  In order to
control the effects of recursion we allow reduction 
of an iteration
only if the
argument is already calculated, i.e., if the argument is a list.

\begin{DEF}[Lists]
\mylabel{D:Lists}
\emph{Lists} $\varL$ (with $\varnatn$ entries) are terms of the form
$$\cons{\varTypeA}\varTermd_1^{\typeD}\varTerma_1^{\varTypeA}
(\ldots(\cons{\varTypeA}\varTermd_n^{\typeD}\varTerma_n^{\varTypeA}
\nil{\varTypeA}))$$ where the $\varTermd_1, \dots, \varTermd_n$ 
stand for arbitrary terms of type $\typeD$.
\end{DEF}

It should be noted that we could also have required the $\linvec{\varTermd}$
to be variables and get the same results. However, this definition allows
more reductions (see Definition~\ref{outer-red}) and therefore is
slightly more flexible when dealing with terms that are not
almost closed, i.e., contain free variables of types other than~$\typeD$.

\begin{DEF}[Conversions]
\mylabel{outer-red}
$\redel$ is defined as:
$$\begin{array}{lcl}
(\lambda\varx^{\varTypeA}.\,\varTermt)\varTerms&\redel&\varTermt\subst{\varTerms}
{\varx} \\
\conPair{\varTermt}{\varTerms}\true&\redel&\varTermt \\
\conPair{\varTermt}{\varTerms}\false&\redel&\varTerms \\
\true\conPair{\varTermt}{\varTerms}&\redel&\varTermt \\
\false\conPair{\varTermt}{\varTerms}&\redel&\varTerms \\
\conTens{\varTypeA}{\varTypeB}\varTermt\varTerms(\lambda\varx^{\varTypeA},
\vary^{\varTypeB}.\,\varTermr)&\redel&\varTermr\subst{\varTermt,\varTerms}
{\varx, \vary} \\
\nil{\varTypeA}\conItA{\varTermt}\varTerms&\redel&\varTerms \\
\cons{\varTypeA}\varTermd^{\typeD}\varTerma\varL\conItA{\varTermt}\varTerms
&\redel&\varTermt\varTermd^{\typeD}\varTerma(\varL\conItA{\varTermt}\varTerms)
\quad\varL\mbox{~a list}
\end{array}$$
\end{DEF}

Although they look quite similar, the rules
$\conPair{\varTermt}{\varTerms}\true\redel\varTermt$ and
$\true\conPair{\varTermt}{\varTerms}\redel\varTermt$ actually have a
very different meaning: The first rule says that we can unfold a
projection once the argument of the term of $\typeCross$-type is in
canonical form, whereas the other rule tells us to take the
if-branch, once we know that the conditional evaluates to $\true$.
Also notice the different typings of the two rules.

\begin{DEF}
\mylabel{def-term-closure}
The reduction relation $\varTermt\red\varTermt^{\prime}$ is inductively defined
as follows:
$$\Rule{\varTermt\redel\varTermt^{\prime}}{\varTermt\red\varTermt^{\prime}}
 \Rule{\varTermt\red\varTermt^{\prime}}{\varTermt\varTerms\red\
                                        \varTermt^{\prime}\varTerms}
 \Rule{\varTerms\red\varTerms^{\prime}}{\varTermt\varTerms\red\
                                        \varTermt\varTerms^{\prime}}$$
\end{DEF}

The requirement that we can only unfold one step of an iteration if
the argument is a full list (in the sense of Definition~\ref{D:Lists})
is crucial for our method of estimating the length of reduction
sequences (cf.\ the introduction to Section~\ref{S:LengthRed}).  This
can be seen from the corresponding case in the proof of
Theorem~\ref{main-theorem}.  Apart from this restriction and the
requirement that one cannot reduce under abstractions the reduction
strategy is arbitrary.

As usual, we call a term $\varTermt$ \emph{normal} if it cannot be further
reduced, i.e., if there is no $\varTermt'$ such that
$\varTermt\red\varTermt'$.

\begin{lem} \mylabel{lem-typ-subst}
			If
            $\typed{\varKon_1\cup\{\varx^{\varTypeB}\}}
				{\varTermt}{\varTypeA}$ and
            $\typed{\varKon_2}{\varTerms}{\varTypeB}$ and moreover
			$\varKon_1$ and $\varKon_2$ are disjoint, then
            $\typed{\varKon_1,\varKon_2}{(\varTermt\subst{\varTerms}{\varx
	        ^{\varTypeB}})} {\varTypeA}$.
\end{lem}
\bbew
\Ind{\varTermt}, using Lemma~\ref{lem-inversion}: If $\varTermt$ is a
variable or a constant the claim is obvious. If $\varTermt$ is of the
form $\varTermt=\varTermt_1\varTermt_2$ then by
Lemma~\ref{lem-inversion} either $\varTermt\subst{\varTerms}{\varx}=
\varTermt_1\subst{\varTerms}{\varx}\varTermt_2$ or
$\varTermt\subst{\varTerms}{\varx} =
\varTermt_1(\varTermt_2\subst{\varTerms}{\varx})$. We apply the
induction hypothesis to the corresponding subterm and can type
$\varTermt\subst{\varTerms}{\varx}$ by the same rule used to type
$\varTermt$. If $\varTermt$ is of the form
$\varTermt=\lambda\vary.\varTermr$, then $\varTermt$ must be typed due
to $(\typeLinTo^+)$, hence
$\typed{\varKon_1\cup\{\varx,\vary\}}{\varTermr}{\varTypeA'}$ and we
may apply the induction hypotheses to $\varTermr$ (without
loss of generality $\vary$ does not to occur in $\varKon_2$) and then
conclude the claim by $(\typeLinTo^-)$. Similar if
$\varTermt=\conPair{\varTermt_1}{\varTermt_2}$. 
\ebew

\begin{lem}[Subject reduction]\mylabel{lem-typ-red}
If $\typed{\varKon}{\varTermt}{\varTypeA}$ and
$\varTermt\red\varTermt^{\prime}$, then
$\typed{\varKon}{\varTermt^{\prime}} {\varTypeA}$.
\end{lem}

\bbew
\Ind{\varTermt\red\varTermt'} 
shows that only conversions have to be considered.
The only non-trivial case is handled in Lemma~\ref{lem-typ-subst}.
\ebew

Next we show that we have sufficiently many reduction rules,
i.e., that normal terms have the expected form. As the size of a term
is its number of free variables, we also have to consider
non-closed terms. As $\typeD$ is the only base type without
closed terms, we can restrict ourselves to \emph{almost closed} terms, i.e.,
to terms with free variables of type $\typeD$ only. It should be noted
that (for example) an arbitrary list can be encoded as an almost closed term.

\begin{prop}\mylabel{prop-numeral}
Every normal, almost closed term of type 
$\typeL{\varTypeA}$, $\typeBool$ or $\typeD$ is
a list, $\true$ or $\false$, or a variable of type $\typeD$,
respectively.
\end{prop}

\bbew
\Ind{\varTermt} and case distinction according to Lemma~\ref{lem-kopfform}:
If $\varTermt$ is of the form $\varx\linvec{\varTermt}$ then $\varx$ has
to be of type $\typeD$ and hence $\vec{\varTermt}$ has to be
empty (since there is no elimination rule for the type $\typeD$). 
If $\varTermt$ is of the form
$(\lambda\varx.\varTerms)\linvec{\varTermt}$ or
$\conPair{\varTerms}{\varTermr}\vec{\varTermt}$, then $\vec{\varTermt}$
has to be empty as well, for otherwise the term would not be normal.
So the interesting case is if $\varTermt$ is of the form
$\varCon\linvec{\varTermt}$. In this case we distinguish cases according to
$\varCon$. If $\varCon$ is $\true$ or $\false$, $\nil{\varTypeA}$ or
$\conTens{\varTypeA}{\varTypeB}$, then $\linvec{\varTermt}$ can
only consist of at most $0$, $1$ or $2$ terms, respectively (for
otherwise there were a redex); hence we have the claim (or the
term is not of one of the types we consider). In the case
$\cons{\varTypeA}\varTermd \varTerma\varTermr\linvec{\varTermt}$ we apply the
induction hypothesis to $\varTermr$ yielding that $\varTermt$ is of
the form $\varL\linvec{\varTermt}$ with a \emph{list} $\varL$. So
if $\linvec{\varTermt}$ would consist of more then one term there
would be a redex; but if $\linvec{\varTermt}$ is a single term, then
the whole term would have arrow-type.
\ebew

It should be noted that in the above proof the only place where
the induction hypothesis actually
is needed is to argue that in $\cons{\varTypeA}
\varTermd^{\typeD}\varTerma \varTermr$ the subterm $\varTermr$ has to be a
list as well (and hence the whole term is a list). 
In particular, these statements need not all be proved simultaneously, but
only simultaneously with the statement that every normal, almost
closed term of list-type is in fact a list (which is necessary due to
the side-condition on the reduction rules for lists).
This modularity is a feature of the
vector notation, i.e., the consistent use of
elimination-\emph{rules} written as \emph{applications}: 
in this way
the (syntactical) Lemma~\ref{lem-kopfform} gives easy access
to the ``canonical'' redex by showing the corresponding introduction. 
It should be noted that the use of elimination-\emph{constants}
(``iterators'') would have messed up this modularity, as in the case
of an elimination constant followed by some arguments one would in
fact need some form of induction hypothesis stating that the argument
to be eliminated is introduced canonically.
\end{section}

\begin{section}{Length of reduction chains}
\mylabel{S:LengthRed}
Before continuing we give a sketch of the main idea of the proof in order to
motivate the following definitions. Since the system is linear,
$\beta$-reduction reduces the number of symbols (since the
$\lambda$ disappears). The same is true for the reductions due to projections
and the ``if\dots then\dots else''.
So the only case where the number of symbols is increased is the case of
iteration. However, if we unfold an iteration \emph{completely}, i.e., if we
reduce
$\cons{\varTypeA}\varTermd_{1}^{\typeD}\varTermr_1^{\varTypeA}
(\ldots(\cons{\varTypeA}\varTermd_{\varnatn}^{\typeD}\varTermr_{\varnatn}^
{\varTypeA}\nil{\varTypeA})) 
\conItA{\varTermh}\varTermt$
in $\varnatn+1$ steps to
$\varTermh\varTermd_1^{\typeD}\varTermr_1^{\varTypeA}
(\ldots(\varTermh\varTermd_{\varnatn}^{\typeD}\varTermr_{\varnatn}^
{\varTypeA}\varTermt))$ then the $\conItA{\cdot}$'s disappear! 
So by making them
``sufficiently heavy'' the total weight of the term reduces. As however the
weight necessary for the $\conItA{\cdot}$-construct depends not only on the
term within the braces, but also on the length $\varnatn$ of the numeral we
iterate on, we cannot assign this term a fixed number; instead we assign a
polynomial(ly bound function)
in Definition~\ref{def-pollen}. But what to plug in? Noting that
we have $\varnatn$ terms $\varTermd_1^{\typeD},\ldots,\varTermd_{\varnatn}
^{\typeD}$ of type $\typeD$ and remembering the idea that terms of type
$\typeD$ should contain free variables (as will be proved in 
Proposition~\ref{lem-term-dia}) we find that the number of free variables
is just the right upper bound for the length $\varnatn$ 
(since the $\linvec{\varTermd}$ are applied
to one another, their free variables have to be disjoint). 

This would lead to
a proof that every reduction strategy
that {\em always unfolds iterations completely} is polynomially bounded. In
order to get results for \emph{every} reduction strategy we notice that
within the subterm 
$\cons{\varTypeA}\varTermd_{1}^{\typeD}\varTermr_1^{\varTypeA}
(\ldots(\cons{\varTypeA}\varTermd_{\varnatn}^{\typeD}\varTermr_{\varnatn}^
{\varTypeA}\nil{\varTypeA})) 
\conItA{\varTermh}\varTermt$ the iteration unfolds at most $\varnatn$ times,
even if the actual number of free variables in the whole term is larger%
\footnote{
This subterm might not unfold at all if it is positioned under a 
$\lambda$-abstraction that remains in the normal form. We do not allow to
reduce under $\lambda$-abstractions (see Definition~\ref{def-term-closure})
since the number of free variables under a $\lambda$-abstraction is 
potentially higher and therefore numerals and lists are potentially longer.
This restriction corresponds to not allowing the use of a ``potential
resource'', i.e., a resource that is not yet present.
}. Using this information we can limit the assigned 
polynomial(ly bounded function) to get a measure that actually decreases in
every step (Theorem~\ref{main-theorem}).

So we use three different measures for terms: the number of free variables, 
which
corresponds to the ``size function'' in Hofmann's
work~\cite{Hofmann99}; the length,
which is the number of symbols of a term that can be accessed (using the
interpretation that you can only access one component of an ordinary pair)
and the polynomial which is the upper bound for the complexity of a function.

\begin{DEF}[Length]\mylabel{def-len}
The length $\len{\varTermt}$ of a term $\varTermt$ 
is inductively defined as follows:
\begin{alignat*}{2}
&\len{\varCon} := \len{\varx}&&:=1
\\
&\len{\varTermt\varTerms}&&:=\len{\varTermt}+\len{\varTerms} 
\\
&\len{\lambda\varx^{\varTypeA}.\,\varTerms}&&:=\len{\varTerms}+1 
\\
&\len{\conPair{\varTermt}{\varTerms}}&&:=
\max{\len{\varTermt},\len{\varTerms}}+1 
\\
&\len{\conItA{\varTermt}}&&:=0
\end{alignat*}
\end{DEF}

As the length $\len{\varTermt}$ is essentially used to handle
$\beta$-redexes, the length of $\conItA{\cdot}$-terms (which are
closed terms!) is of no importance. So for simplicity, and to obtain
slightly sharper results (as the length occurs in the
Definition~\ref{def-pollen} of the polynomial bound) the value $0$ has
been chosen. It should be noted that all the results would also hold
with any other ``reasonable'' definition for
$\len{\conItA{\varTermt}}$, such as $\len{\varTermt}$.

\begin{lem}\mylabel{lem-len-sub}
If $\typed{\varKon_1}{\varTermt}{\varTypeA}$ and
$\typed{\varKon_2}{\varTerms}{\varTypeB}$ and moreover
$\varKon_1,\varKon_2$ are disjoint, then
$\len{\varTermt\subst{\varTerms}{\varx}}\leq\len{\varTermt}
+\len{\varTerms}$.
In particular, $\len{(\lambda x^{\varTypeA}.\,\varTermt)\varTerms} >
                  \len{\varTermt\subst{\varTerms}{\varx}}$.
\end{lem}
\bbew
\Ind{\varTermt}, using the fact that $\varTermt$ is typed and therefore in the
case of an application only one of the terms can contain the 
variable~$\varx$ free (compare Lemma~\ref{lem-inversion}).

For instance, if $\varTermt$ is $\varTermt_1 \varTermt_2$, then the
last rule of a derivation of $\typed{\varKon}{\varTermt_1,
\varTermt_2}{\varTypeA}$ must be one of $(\typeLinTo^-)$,
$(\typeCross_0^-)$, $(\typeCross_1^-)$, $(\typeBool^-)$,
$(\typeTensor^-)$ or $(\typeL{\varTypeA}^-)$. In each of these cases
the claim is obvious.
\ebew

\begin{lem}
\mylabel{lem-term-dia} Every term of type $\typeD$ contains a free
variable, i.e., if
$\typed{\varKon}{\varTermt}{\typeD}$, then $\varKon\neq\emptyset$.
\end{lem}
\bbew
\Ind{\len{\varTermt}} and case distinction according to 
Lemma~\ref{lem-kopfform}. 
The case $\varx\linvec{\varTermt}$ is trivial. In the case
$(\lambda\varx.\varTerms)\linvec{\varTermt}$ the $\linvec{\varTermt}$
cannot be empty (for otherwise the term would have an arrow type), so we
can apply the induction hypothesis to the (by Lemma~\ref{lem-len-sub})
shorter term
$\varTermt\subst{\varTermt_1}{\varx}\varTermt_2\ldots\varTermt_{\varnatn}$.
Similarly for $\conPair{\varTermr}{\varTerms}\linvec{\varTermt}$,
where we apply the induction hypothesis to 
$\varTermr\varTermt_2\ldots\varTermt_{\varnatn}$ or
$\varTerms\varTermt_2\ldots\varTermt_{\varnatn}$ depending whether
$\varTermt_1$ is $\true$ or $\false$ (note that by \emph{typing} one
of these has to be the case) and 
$\conTens{\varTypeA}{\varTypeB}\linvec{\varTermt}$, where we again use
Lemma~\ref{lem-len-sub}.
For $\cons{\varTypeA}\linvec{\varTermt}$ we use the induction hypothesis for 
$\varTermt_1$. For $\nil{\varTypeA}\linvec{\varTermt}$
use the induction hypothesis for $\varTermt_2\ldots\varTermt_{\varnatn}$.
In the cases $\true\linvec{\varTermt}$ and $\false\linvec{\varTermt}$,
by typing, $\varTermt_1$ has to be of the form
$\conPair{\varTermr}{\varTerms}$ and we can apply the induction
hypothesis to (say) $\varTermr\varTermt_2\ldots\varTermt_{\varnatn}$.
\ebew

The following corollary is formulated for type $1$ functions only.
However, if one accepts the number of free variables as a reasonable
size measure
for definable functions for higher types as well, then the corollary trivially
remains valid for \emph{every} definable function.

\begin{cor}[Non-size-increasing pro\-per\-ty]\mylabel{cor-non-size}
Every function of type $\typeL{\varTypeA} \typeLinTo \typeL{\varTypeA}$
definable by a closed term has the property that
the output is not longer than the input.
\end{cor}

\bbew Lemma~\ref{lem-term-dia} shows that for closed terms of type
$\typeL{\varTypeA}$ the usual length and the number of free variables
coincide (due to the typing $\typeD \typeLinTo \varTypeA \typeLinTo
\typeL{\varTypeA} \typeLinTo \typeL{\varTypeA}$ of the 
$\cons{\varTypeA}$ function).
The number of free variables trivially does not increase when reducing the
term to normal form.
\ebew

Let $\N$ be the set of natural numbers and
$\Np$ be the set of all functions from $\N$ to $\N$
that are (pointwise) bounded by some polynomial. We write $\X$ for the
identity on the natural numbers, $\varpolf+\varpolg$ and $\varpolf
\cdot\varpolg$ for the pointwise sum and product in $\Np$ 
and $\X_{\varnatn}$ for the minimum of
the identity and $\varnatn$, i.e., $\X_{\varnatn}(\varnatm)=\min{\varnatn,
\varnatm}$. $\N$ is treated as a subset of $\Np$ by identifying a natural
number with the corresponding constant function. Let $\polleq$ be the
pointwise order on $\Np$. Then there exist finite suprema (the pointwise
maxima) in $\Np$ with respect to $\polleq$, denoted by $\suppol{\cdot}$.

It seems that by considering functions we leave the realm of syntax. But
as we will only use functions that are explicitly defined by 
pointwise sum, product, maximum and minimum from the identity and constants 
we could as well restrict ourselves to these functions only. It should be
obvious how these functions can be encoded by some finite 
(syntactical) object.

\begin{DEF}[Polynomial bound $\pollen{\cdot}$ of a term]
\mylabel{def-pollen}
For each (not necessarily typed) term $\varTermt$ we define
a polynomial
$\pollen{\varTermt}\in\Np$ by recursion on $\varTermt$.
\begin{alignat*}{2}
&\pollen{\varx} := \pollen{\varCon} &&:= 0
\\
&\pollen{\varTermt\varTerms} &&:= 
\begin{cases}
\pollen{\varTermt} +
\X_{\varnatn}\cdot\pollen{\varTermh}
+\X_{\varnatn}\cdot\len{\varTermh}
& \text{if $\varTermt$ is a list
with $\varnatn$ entries}
\\ 
& \text{and $\varTerms$ is of the form $\conItA{\varTermh}$}
\\[1ex]
\pollen{\varTermt}+\pollen{\varTerms} & \text{otherwise}
\end{cases}
\\
&\pollen{\lambda\varx^{\varTypeA}.\,\varTermt}  &&:= \pollen{\varTermt} 
\\
&\pollen{\conPair{\varTermt}{\varTerms}}
 &&:= \suppol{\pollen{\varTermt}, \pollen{\varTerms}} 
\\
&\pollen{\conItA{\varTermh}}
 &&:= \X\cdot\pollen{\varTermh}+\X\cdot\len{\varTermh}
\end{alignat*}
We write $\pollen{\linvec{\varTermt}}$ for
$\sum\limits_{\varnati=1}^{\varnatn}\pollen{\varTermt_{\varnati}}$.
\end{DEF}

Immediately from the definition (since $\X_{\varnatn}\polleq\X$) we get
\begin{prop}\mylabel{prop-pol-app}
$\pollen{\varTermt\varTerms}\polleq\pollen{\varTermt}+\pollen{\varTerms}$.
\hfill \qedsymbol
\end{prop}

\begin{lem}
\mylabel{lem-pol-sub}
If $\typed{\varKon_1}{\varTermt}{\varTypeA}$, $\typed{\varKon_2}{\varTerms}
{\varTypeB}$ and $\varKon_1,\varKon_2$ are disjoint, then
$\pollen{\varTermt\subst{\varTerms}{\varx}}\polleq
\pollen{\varTermt}+\pollen{\varTerms}$.
\end{lem}
\bbew We prove this by induction on $\varTermt$. The only non-trivial case
is when $\varTermt$ is an application (note that the case $\conItA{\varTermh}$
does not occur since $\varTermt$ is
typed). So let $\varTermt$ be of the form $\varTermr\varTermr^{\prime}$.
We distinguish cases whether $\varTermr$ is a list or not.

If $\varTermr$ is a list with $\varnatn$ entries then, 
as $\varTermt$ is typed,
we know that $\varTermr^{\prime}$ must be of the form $\varTermr^{\prime}
= \conItA{\varTermh}$ and therefore closed. 
$\varTermr\subst{\varTerms}{\varx}$ is easiliy seen to be
again a list with $\varnatn$ entries. 
Therefore we get
\begin{align*}
\pollen{(\varTermr\varTermr^{\prime})\subst{\varTerms}{\varx}} 
&= \pollen{\varTermr\subst{\varTerms}{\varx}\varTermr^{\prime}}
\\
&= \pollen{\varTermr\subst{ \varTerms}{\varx}}+
 \X_{\varnatn}\cdot\pollen{\varTermh} +\X_{\varnatn}\cdot \len{\varTermh}
\\
&\polleq 
\pollen{\varTermr}+\pollen{\varTerms}+
\X_{\varnatn}\cdot\pollen{\varTermh}
+\X_{\varnatn}\cdot\len{\varTermh}
\quad \text{by IH}
\\
&= \pollen{\varTermr\varTermr^{\prime}} + \pollen{\varTerms}.
\end{align*}
If $\varTermr$ is not a list, then from the fact that
$\varTermt$ is typed we know that at most one of the terms
$\varTermr$ and $\varTermr^{\prime}$ contains the
variable $\varx$ free. In the case $\varx\in\FV{\varTermr}$ (the other case
is handled similarly) we have
\begin{align*}
\pollen{(\varTermr\varTermr^{\prime})\subst{\varTerms}{\varx}}
& = \pollen{\varTermr\subst{\varTerms}{\varx}\varTermr^{\prime}}
\\
&\polleq 
\pollen{\varTermr\subst{\varTerms}{\varx}}+\pollen{\varTermr^{\prime}}
\quad \text{by Proposition~\ref{prop-pol-app}}
\\
&\polleq 
\pollen{\varTermr}+\pollen{
 \varTerms}+\pollen{\varTermr^{\prime}}
\quad \text{by IH}
\\
& =\pollen{\varTermr\varTermr^{\prime}}
 +\pollen{\varTerms}.
\end{align*}
This completes the proof.
\ebew

\begin{satz}\mylabel{main-theorem}
Assume $\typed{\varKon}{\varTermt}{\varTypeA}$ and
$\varnatN\geq\card{\FV{\varTermt}}$ where $\varnatN\in\N$. If
$\varTermt\red\varTermt^{\prime}$, then
$$\pollen{\varTermt}(\varnatN)+\len{\varTermt} >
  \pollen{\varTermt^{\prime}}(\varnatN)+\len{\varTermt^{\prime}}$$
In particular, \emph{any} reduction sequence starting from $\varTermt$ has
length at most $$\pollen{\varTermt}(\card{\FV{\varTermt}})+\len{\varTermt}$$
\end{satz}

\bbew
We prove this by induction on the definition of the relation $\varTermt
\red\varTermt^{\prime}$.

\textbf{Case} $\varTermr\varTerms\red\varTermr^{\prime}\varTerms$ via
$\varTermr\red\varTermr^{\prime}$.   We distinguish whether $r$ is a
list or not.

\emph{Subcase} $\varTermr$ is a list with $\varnatn$ entries.
Then by typing restrictions and from Lemma~\ref{lem-term-dia} we know that
$\varnatn\leq\card{\FV{\varTermt}}\leq\varnatN$. Also 
$\varTerms$ has to be of the form $\varTerms=\conItA{\varTermh}$.
$\varTermr^{\prime}$ is again a list with $\varnatn$ entries, so
$$
\begin{array}{clcl}
&\pollen{\varTermr\varTerms}(\varnatN)&+&\len{\varTermr\varTerms}\\
=&
\pollen{\varTermr}(\varnatN)+
\varnatn\cdot\pollen{\varTermh}(\varnatN)
 +\varnatn\cdot\len{\varTermh}
&+&\len{\varTermr}+\len{\varTerms}
\\ 
>&
\pollen{\varTermr^{\prime}}(\varnatN)+
 \varnatn\cdot{\pollen{\varTermh}}(\varnatN)
 +\varnatn\cdot{\len{\varTermh}}
&+&\len{\varTermr^{\prime}}+\len{\varTerms} 
\quad\hbox{by IH}
\\
= &\pollen{\varTermr^{\prime}\varTerms}(\varnatN)&+&\len{\varTermr^{\prime}
\varTerms}
\end{array}
$$
\emph{Subcase} $\varTermr$ is not a list.   Then
$$
\begin{array}{clcl}
&\pollen{\varTermr\varTerms}(\varnatN)&+&\len{\varTermr\varTerms}
\\ 
=&
\pollen{\varTermr}(\varnatN)+\pollen{\varTerms}(\varnatN)&+&\len{\varTermr}
+\len{\varTerms} 
\\
>&
\pollen{\varTermr^{\prime}}(\varnatN)+\pollen{\varTerms}(\varnatN)&+&
\len{\varTermr^{\prime}} +\len{\varTerms} \quad\hbox{by IH}
\\
\geq &
\pollen{\varTermr^{\prime}\varTerms}(\varnatN)&+&\len{\varTermr^{\prime}
\varTerms}
\quad\hbox{by Proposition~\ref{prop-pol-app}.}
\end{array}
$$

\textbf{Case} $\varTermr\varTerms\red\varTermr\varTerms^{\prime}$ via
$\varTerms\red\varTerms^{\prime}$.   Then $\varTermr$ is not
a list, since otherwise, $\varTerms$ were of the form
$\conItA{\cdot}$ and we must not reduce within
braces. Therefore this case can be handled as the second subcase above.

\textbf{Case} $\varTermt\redel\varTermt^{\prime}$.   We distinguish
subcases according to the form of the conversion.

\emph{Subcase} $(\lambda\varx.\,\varTermt)\varTerms
\redel\varTermt\subst{\varTerms}{\varx}$.   We have
$$
\begin{array}{clcl}
&\pollen{(\lambda\varx.\,\varTermt)\varTerms}(\varnatN)&+&
  \len{(\lambda\varx.\,\varTermt)\varTerms}
\\ 
=&
\pollen{\varTermt}(\varnatN)+\pollen{\varTerms}(\varnatN)&+&
\len{\varTermt}+1+\len{\varTerms} 
\\
\geq&
\pollen{\varTermt\subst{\varTerms}{\varx}}(\varnatN) &+&
\len{\varTermt}+1+\len{\varTerms}
\quad\hbox{by Lemma~\ref{lem-pol-sub}.}
\end{array}
$$
\emph{Subcase}
$\conTens{\varTypeA}{\varTypeB}\varTermr\varTerms(\lambda\varx,\vary.
\varTermt) \redel\varTermt\subst{\varTermr,\varTerms}{\varx,\vary}$]
is handled similarly.

\emph{Subcase} 
$\cons{\varTypeA}\varTermd^{\typeD}\varTerma\varL\conItA{\varTerms}\varTermr
\redel\varTerms\varTermd^{\typeD}\varTerma(\varL\conItA{\varTerms}\varTermr)
$] 
with $\varL$ a list with $\varnatn$ entries.
Then by Lemma~\ref{lem-term-dia} and the linear typing discipline
$\varnatn+1\leq\card{\FV{\cons{\varTypeA}\varTermd^{\typeD}\varTerma
\varL\conItA{\varTerms}\varTermr}}\leq\varnatN$. Therefore we have
$$
\begin{array}{ccl}
&& \pollen{\cons{\varTypeA}\varTermd^{\typeD}\varTerma\varL
  \conItA{\varTerms}\varTermr}(\varnatN) 
\\&+& \len{\cons{\varTypeA}\varTermd^{\typeD}\varTerma\varL
  \conItA{\varTerms}\varTermr}
\\ \\
=&&\pollen{\varTermd}(\varnatN)+
  \pollen{\varTerma}(\varnatN)+\pollen{\varL}(\varnatN)
\\&&
  {}+(n+1)\pollen{\varTerms}(\varnatN)+(n+1)\len{\varTerms}+ \pollen{\varTermr}
  (\varnatN)
\\ 
&+& 1 + \len{\varTermd}+\len{\varTerma}+\len{\varL}+\len{\varTermr}
\end{array}
 $$
and on the other hand
$$
\begin{array}{ccl}
&&\pollen{
\varTerms\varTermd^{\typeD}\varTerma(\varL\conItA{\varTerms}\varTermr)}
(\varnatN)\\
&+&
\len{\varTerms\varTermd^{\typeD}\varTerma(\varL\conItA{\varTerms}\varTermr)}
\\\\=&&
\pollen{\varTerms}(\varnatN) +\pollen{\varTermd}(\varnatN)
+\pollen{\varTerma}(\varnatN)  
+\pollen{\varL}(\varnatN) \\&&
{}+\varnatn\cdot\pollen{\varTerms}(\varnatN)
+\varnatn\cdot\len{\varTerms} 
+\pollen{\varTermr}(\varnatN)
\\&+&
\len{\varTerms}+\len{\varTermd}+\len{\varTerma}+\len{\varL}
+\len{\varTermr}
\end{array}
$$
which obviously is strictly smaller.

\emph{Subcase} $\true\conPair{\varTermr}{\varTerms}\redel\varTermr$.  
We have
$$\begin{array}{clcl}
&\pollen{\true\conPair{\varTermr}{\varTerms}}(\varnatN) &+&
 \len{\true\conPair{\varTermr}{\varTerms}} \\
=&\suppol{\pollen{\varTermr},\pollen{\varTerms}}(\varnatN) &+&
 1 + \max{\len{\varTermr},\len{\varTerms}} \\
\geq&\pollen{\varTermr}(\varnatN) &+& 1 + \len{\varTermr}
\end{array}$$
The remaining subcases are similar to the last one.
\ebew

\begin{cor}[Hofmann, 1999]
If $\typed{\emptyset}{\varTermt}{\typeL{\typeBool} \typeLinTo \cdots
\typeLinTo \typeL{\typeBool} \typeLinTo \typeL{\typeBool}}$,
then $\varTermt$ denotes a function computable in polynomial many
steps.
\end{cor}

\bbew We only treat unary functions.  Let $\varL$ be an almost closed
normal list.  Then $\pollen{\varTermt\varL} =\pollen{\varTermt}$, as
$\pollen{\varL}=0$.  Therefore an upper bound for the number of steps
is
$\pollen{\varTermt}(\card{\FV{\varL}})+\len{\varTermt}+\len{\varL}$,
which is polynomial in the length of the input $\varL$.  \ebew

Strictly speaking, the above result does not show that functions
definable by closed terms are polynomial time computable, since no
machine model has been provided.
However, it is rather obvious how these reductions can be
implemented in constant time (with respect to the size of the input): 
First note that the only terms that have to be duplicated are the
step terms of iterations which are \emph{not} part of the input
and ---~as they are closed terms~--- no part of the input can become
substituted into them.
Next remark that all $\beta$-redexes are linear and hence can be
implemented by the usual pointer switching techniques. In the case of
$\conPair{\cdot}{\cdot}$ an indirection node is
used for the shared variables of the two components. Our reduction
strategy does not
reduce within a pair until the projection or the
``if\dots then\dots else''-statement is resolved. As pairs are
not part of the input, the amount of extra work in each resolution
step is independent of the input.

\begin{bsp}[Insertion sort]\mylabel{bsp-insert}
Let $\varTypeA$ be a type equipped with a linear ordering. Assume
that this ordering is represented by the term $\leqbsp{\varTypeA}$ of
type $ \varTypeA\typeLinTo
\varTypeA\typeLinTo\typeBool\typeTensor(\varTypeA\typeTensor\varTypeA)$,
i.e., 
$\leqbsp{\varTypeA}\varTermt\varTerms\red^{\ast}\conTens{\typeBool}
{\varTypeA\typeTensor\varTypeA}\true(\conTens{\varTypeA}{\varTypeA}\varTermt
\varTerms)$, if $\varTermt$ is ``smaller'' than $\varTerms$, and
$\leqbsp{\varTypeA}\varTermt\varTerms\red^{\ast}\conTens{\typeBool}
{\varTypeA\typeTensor\varTypeA}\false(\conTens{\varTypeA}{\varTypeA}\varTermt
\varTerms)$ otherwise.

Using this function, we can define a sort function 
of type $\varTypeA\typeLinTo\varTypeA\typeLinTo\varTypeA\typeTensor\varTypeA$
for two elements, i.e., 
a function of two arguments that returns them in the correct order:
\[
\leqbsp{\varTypeA}^{\prime}=\lambda\varp_1^{\varTypeA},\varp_2^
{\varTypeA}.\,\leqbsp{\varTypeA}\varp_1\varp_2 (\lambda\vary^{\typeBool},
\varp^{\varTypeA\typeTensor\varTypeA}.\,\varp
(\lambda\varp_1^{\varTypeA},\varp_2^{\varTypeA}.
\vary\conPair{\conTens{\varTypeA}{\varTypeA}\varp_1\varp_2}
             {\conTens{\varTypeA}{\varTypeA}\varp_2\varp_1})
\]
Now we can immediately define a function of type
$\typeL{\varTypeA}\typeLinTo\typeD\typeLinTo\varTypeA\typeLinTo
\typeL{\varTypeA}$
inserting an element at the correct position in a given sorted list.
At each step we
compare the given element with the first element of the list, put the
smaller element at the beginning of the list and insert the larger
one in the rest of the list.
\begin{align*}
{\tt insert} = \lambda\varl.&\varl\{\lambda\varx_1^{\typeD},\vary_1^
{\varTypeA},\varp^{\typeD\typeLinTo\varTypeA\typeLinTo\typeL{\varTypeA}},
\varx_2^{\typeD},\vary_2^{\varTypeA}.
\leqbsp{\varTypeA}^{\prime}\vary_1\vary_2(\lambda\varz_1^{\varTypeA},\varz_2
^{\varTypeA}\,\cons{\varTypeA}\varx_1\varz_1(\varp\varx_2\varz_2))\}
\\ &
(\lambda\varx^{\typeD},\vary^{\varTypeA}.\,\cons{\varTypeA}\varx\vary\nil{
\varTypeA})
\end{align*}
Then insertion sort is defined as usual:
$$
\mathtt{sort}=\lambda\varl^{\varTypeA}.\,
\varl\conItA{\lambda\varx^{\typeD},\vary^{\varTypeA},
\varl^{\typeL{\varTypeA}}.\,
\mathtt{insert}\varl\varx\vary}
\nil{\varTypeA}
$$

Counting braces, as $\pollen{\leqbsp{\varTypeA}
 ^{\prime}}=\pollen{\leqbsp{\varTypeA}}$,  we get
$$\pollen{\mathtt{sort}}=\X\cdot\pollen{\mathtt{insert}}
+O(\X)=\X^2\cdot\pollen{\leqbsp{\varTypeA}^{\prime}}+O(\X^2)
=\X^2\cdot\pollen{\leqbsp{\varTypeA}}+O(\X^2)$$ 
This reflects the fact that insertion sort is quadratic in the
number of comparison operations.
\end{bsp}

By simple modifications of (the proof of)~\cite[\S 4.3]{Hofmann99}
we may conclude that many ``natural orderings'', e.g., the normal
ordering on binary coded natural numbers, \emph{can} be defined in the  given
term system.

However,
it should be noted that it is \emph{not necessary} that every interesting
ordering is definable in the given system. It would also be possible to
just add a new symbol $\leqbsp{\varTypeA}$ with the conversion rules
$\leqbsp{\varTypeA}\varTermt\varTerms\redel\conTens{\typeBool}
{\varTypeA\typeTensor\varTypeA}\true(\conTens{\varTypeA}{\varTypeA}\varTermt
\varTerms)$, if $\varTermt$ is ``smaller'' than $\varTerms$, and
$\leqbsp{\varTypeA}\varTermt\varTerms\redel\conTens{\typeBool}
{\varTypeA\typeTensor\varTypeA}\false(\conTens{\varTypeA}{\varTypeA}\varTermt
\varTerms)$ otherwise. With $\len{\leqbsp{\varTypeA}}:=4$ the
above theory remains valid and shows that there are at most
$\X^2$ of the newly introduced conversions in a normalizing
sequence. Therefore this theory can be used
to calculate the number of calls to a ``subroutine'', even if the subroutine
itself is not definable in the given system, or not even polynomial time
computable.
 \end{section}

\begin{section}{Extensions of the System}
\mylabel{S:Extensions}
The syntactical analysis of the system allows various extensions which
we only sketch here, giving sufficient detail to reconstruct the
proofs.

\begin{subsection}{Full Polynomial Time}\mylabel{subsec-full-ptime}
The system so far only contains non-size-increasing functions, and
hence cannot contain all \texttt{Ptime} functions.
New results of
Hofmann~\cite{Hofmann01} show that indeed all 
these functions (and hence in particular all {\tt
Ptime}-predicates) are already \emph{within} the present system.
 
Here we shall briefly sketch an approach to obtain all \texttt{Ptime}
functions, that might give
some insight into the way the restriction to non-size-increasing
functions works.  Its motivation was to
avoid explosion of growth by iterating over 
already aggregated data-structures.
Yet in the definition of {\tt Ptime}, 
the only large data-structure of a Turing machine is its
tape. Moreover, a Turing machine does not iterate over its tape but
instead modifies it locally.

The central idea lies in the observation that 
size is represented by the number of free variables. Hence,
we can add a type $\typeI$ that allows closed terms for objects
that are \emph{semantically} of arbitrary size. On this type, we can
then define functions that are \emph{semantically} size-increasing,
like the extension of the Turing-tape, but are \emph{from a technical
point of view} non-size-increasing, in that they do not require an
argument of type $\typeD$.

Iteration on this new type $\typeI$ would lead
beyond polynomial time, as
the number of iterations that a loop (i.e., an
$\typeL{\varTypeA}$-elimination) unfolds to is no longer immediately
related to the input (via the number of free variables).

The term system is extended by
adding a new ground type $\typeI$,
adding the following constants with their respective types
$$\begin{array}{llcll}
\Izero&\typeI&\qquad\qquad&
\Ipred\qquad&\typeI\typeLinTo\typeI\\
\IsucA&\typeI\typeLinTo\typeI&&
\Iiszero\qquad&\typeI\typeLinTo(\typeBool\typeTensor\typeI)\\
\IsucB&\typeI\typeLinTo\typeI&&
\Ifirstdigit\qquad&\typeI\typeLinTo(\typeBool\typeTensor\typeI)
\end{array}$$
$\redel$ is enriched by
$$\begin{array}{lclclcl}
\Ipred\Izero&\redel&\Izero&\qquad\qquad&
\Iiszero\Izero&\redel&\conTens{\typeBool}{\typeI}\true\Izero
\\
\Ipred(\Isuc{\varnati}\varTermt^{\typeI})&\redel&\varTermt^{\typeI}&&
\Iiszero(\Isuc{\varnati}\varTermt^{\typeI})
&\redel&\conTens{\typeBool}{\typeI}\false(\Isuc{\varnati}\varTermt^{\typeI})
\end{array}$$
$$\begin{array}{lcl}
\Ifirstdigit\Izero&\redel&\conTens{\typeBool}{\typeI}\false\Izero\\
\Ifirstdigit(\IsucA\varTermt^{\typeI})&\redel&
\conTens{\typeBool}{\typeI}\false(\IsucA\varTermt^{\typeI})\\
\Ifirstdigit(\IsucB\varTermt^{\typeI})&\redel&
\conTens{\typeBool}{\typeI}\true(\IsucB\varTermt^{\typeI})
\end{array}$$

The definitions of the relations
$\typed{\varKon}{\varTermt}{\varTypeA}$ and $\varTermt\red\varTermt^{\prime}$
remain unchanged. 
Proposition~\ref{prop-term-weak}, and Lemmata~\ref{lem-term-typ},
\ref{lem-kopfform}, \ref{lem-typ-subst} and~\ref{lem-typ-red} remain
valid with almost identical proofs.
 
We call terms of the form
$\Isuc{\varnati_1}(\ldots(\Isuc{\varnati_{\varnatn}}\Izero))$
\emph{short numerals}. As in Proposition~\ref{prop-numeral}
we show that every normal,
almost closed term of type $\typeI$ is a short numeral (and therefore closed).

The definition of the term length $\len{\varTermt}$ had the property
that for all reductions except iteration the length decreased.
To retain this property we define
$\len{\Iiszero}=\len{\Ifirstdigit}=3$
and keep the rest of Definition~\ref{def-len}. In particular, every constant
different to $\Iiszero$ and $\Ifirstdigit$ still has length $1$. Then
Lemmata~\ref{lem-len-sub} and \ref{lem-term-dia} remain valid.

The definition of $\pollen{\varTermt}$ remains unchanged. Then (with
identical proof) Lemma~\ref{lem-pol-sub} holds and also the
main Theorem~\ref{main-theorem}.
In particular, the extended system still consists of {\tt Ptime} functions
only.

To show that every {\tt Ptime} function can be defined by a closed
term of type $\typeN\typeLinTo\typeI$, we code the configuration of a
Turing machine (with $\varnatN$ states $\{S_0,\ldots,S_{\varnatN-1}\}$,
working over the alphabet $\{0,1\}$) 
with the symbols
$\varnati_0\ldots\varnati_{\varnatk}$ before and including the head
and the symbols
$\varnatj_0\ldots\varnatj_{\varnatk^{\prime}}$ followed by the non-visited
positions after the head and with current state
$S_{\varnatm}$ by
\begin{align*}
\typeTensor&(\Isuc{\varnati_{\varnatk}}(\ldots(\Isuc{\varnati_0}\Izero)))\\
  (\typeTensor&(\Isuc{\varnatj_0}(\ldots(\Isuc{\varnatj_{\varnatk^{\prime}}}
  \Izero))) \\
   (\typeTensor&
  \varTermt_{1}(\typeTensor\ldots(\typeTensor\varTermt_{\varnatn-1}\varTermt_
  {\varnatn}))))
\end{align*}
with $\varnatn$ being the
smallest number such that $\varnatN\leq2^{\varnatn}$ and
each of the $\linvec{\varTermt}$ being $\true$ or $\false$ such
that this is the binary coding of $\varnatm$ . 
The closed terms $\IsucA$ and $\IsucB$ extend the Turing
tape where necessary, so we can define the one-step function of the Turing
machine by sufficiently many case-distinctions. Iterating this
one-step function polynomially many times (e.g., as shown by
Hofmann~\cite[\S4.3]{Hofmann99}) completes the (sketched) proof.
\end{subsection}

\begin{subsection}{Trees}
\mylabel{SS:Trees}
We sketch how our technique applies to the more complex data
type of binary labeled trees; this is to be hoped for, as
Hofmann's original method is capable of dealing with 
them~\cite[\S4.3]{Hofmann00}.  Notice that the extension is not
completely obvious, since iteration on trees involves \emph{two}
recursive calls.  It turns out that the
number of free variables in a term still is a good measure for the
number of unfoldings of an iteration.

The system is extended by
\begin{itemize}
\item a new type constructor $\typeT{\varTypeA}{\varTypeB}$ for
trees (with nodes labeled with elements of type $\varTypeA$ and
leaves labeled with elements of type $\varTypeB$)
\item new constants with their respective types
$$
\begin{array}{ll}
\emptyTree{\varTypeA}{\varTypeB}
\qquad&\varTypeB\typeLinTo\typeT{\varTypeA}{\varTypeB} \\
\consTree{\varTypeA}{\varTypeB}\qquad&\typeD\typeLinTo\varTypeA
\typeLinTo\typeT
{\varTypeA}{\varTypeB}
\typeLinTo\typeT{\varTypeA}{\varTypeB}
\typeLinTo\typeT{\varTypeA}{\varTypeB}
\end{array}$$
\item a new term constructor $\conTit{\cdot}{\cdot}$
\item a new typing rule
$$\begin{array}{ll}
\Rule{\typed{\varKon}{\varTermt}{\typeT{\varTypeA}{\varTypeB}}\qquad
      \typed{\emptyset}{\varTerms}
            {\typeD\typeLinTo\varTypeA
			\typeLinTo\varTypeC \typeLinTo\varTypeC \typeLinTo\varTypeC}
	  \qquad
      \typed{\emptyset}{\varTermr}
            {\varTypeB\typeLinTo\varTypeC}
	 }
     {\typed{\varKon}{(\varTermt\conTit{\varTerms}{\varTermr})}
	{\varTypeC}} &
(\typeT{\varTypeA}{\varTypeB}^-)
\end{array}$$
\end{itemize}

We inductively define the notion of a tree (with $\varnatn$ nodes) by
\begin{itemize}
\item $\emptyTree{\varTypeA}{\varTypeB}\varTermt^{\varTypeB}$ is a
tree (with $0$ nodes)
\item if $\varTermt_1$ and $\varTermt_2$ are trees (with $\varnatn_1$
and $\varnatn_2$ nodes) then 
$\consTree{\varTypeA}{\varTypeB}
\varTermd^{\typeD}\varTerma^{\varTypeA}\varTermt_1\varTermt_2$ is also
a tree (with $\varnatn_1+\varnatn_2$ nodes)
\end{itemize}

The conversion relation $\redel$ is augmented by
$$\begin{array}{l}
\emptyTree{\varTypeA}{\varTypeB}\varTermt\conTit{\varTerms}{\varTermr}
\redel\varTermr\varTermt \\
\consTree{\varTypeA}{\varTypeB}\varTermd\varTerma\varTermt_1\varTermt_2
\conTit{\varTerms}{\varTermr}\redel\varTerms\varTermd\varTerma
(\varTermt_1\conTit{\varTerms}{\varTermr})
(\varTermt_2\conTit{\varTerms}{\varTermr})
\qquad\varTermt_1,\varTermt_2\mbox{~trees}
\end{array}$$

We extend the definition of the length by
$$\len{\conTit{\varTerms}{\varTermr}}:=\len{\varTermr}$$
and the definition of the polynomial bound by
\begin{alignat*}{2}
&\pollen{\conTit{\varTerms}{\varTermr}} &&:= 
\X \cdot \pollen{\varTerms} +
(\X+1) \cdot \pollen{\varTermr} +
\X \cdot \len{\varTerms} +
\X \cdot \len{\varTermr} 
\\
&\pollen{\varTermt\varTerms} &&:=
\begin{cases}
\pollen{\varTermt} +
\X_{\varnatn}\pollen{\varTermh}
+\X_{\varnatn}\len{\varTermh}
& \text{if $\varTermt$ is a list with $\varnatn$ entries}
\\ 
& \text{and $\varTerms$ is of the form $\conItA{\varTermh}$}
\\[1ex]
\pollen{\varTermt}+\X_{\varnatn}\cdot\pollen{\varTerms'} +
(\X_{\varnatn}+1)\cdot\pollen{\varTermr'} 
\\
+ \X_{\varnatn}\cdot\len{\varTerms'} +
\X_{\varnatn}\cdot\len{\varTermr'} 
&\text{if $\varTermt$ is a tree with $\varnatn$ nodes}
\\
&\text{and $\varTerms$ is of the form $\conTit{\varTerms'}{\varTermr'}$}
\\[1ex]
\pollen{\varTermt}+\pollen{\varTerms} & \text{otherwise}
\end{cases}
\end{alignat*}
Then the theory in section~\ref{S:LengthRed}
remains valid, with identical proofs.  The new
non-trivial subcase in Theorem~\ref{main-theorem} is
$$
\consTree{\varTypeA}{\varTypeB}\varTermd\varTerma\varTermt_1\varTermt_2
\conTit{\varTerms}{\varTermr}
\redel
\varTerms\varTermd\varTerma
(\varTermt_1\conTit{\varTerms}{\varTermr})
(\varTermt_2\conTit{\varTerms}{\varTermr})
$$
where $\varTermt_1$ and $\varTermt_2$ are trees with $\varnatn_1$ and
$\varnatn_2$ nodes respectively.  Then by Lemma~\ref{lem-term-dia} and
the linear typing discipline we know that
$\varnatn_1 + \varnatn_2 +1 \leq \card{\FV{\consTree{\varTypeA}{\varTypeB}
\varTermd\varTerma\varTermt_1\varTermt_2
\conTit{\varTerms}{\varTermr}}} \leq \varnatN$, and hence

$$
\begin{array}{ccl}
&&\pollen{
\consTree{\varTypeA}{\varTypeB}\varTermd\varTerma\varTermt_1\varTermt_2
\conTit{\varTerms}{\varTermr}}(\varnatN)\\
&+&\len{
\consTree{\varTypeA}{\varTypeB}\varTermd\varTerma\varTermt_1\varTermt_2
\conTit{\varTerms}{\varTermr}}
\\[1ex]
=&&\pollen{\varTermd}(\varnatN)+ \pollen{\varTerma}(\varnatN)+
\pollen{\varTermt_1}(\varnatN)+
\pollen{\varTermt_2}(\varnatN)+\\&+&
(\varnatn_1+\varnatn_2+1)\pollen{\varTerms}(\varnatN)+
(\varnatn_1+\varnatn_2+1+1)\pollen{\varTermr}(\varnatN)\\
&+&(\varnatn_1+\varnatn_2+1)\len{\varTerms}+
(\varnatn_1+\varnatn_2+1)\len{\varTermr}\\
&+& 1 + \len{\varTermd}+\len{\varTerma}+\len{\varTermt_1}+\len{\varTermt_2}
+\len{\varTermr}
\end{array}
$$
and
$$
\begin{array}{ccl}
&&\pollen{
\varTerms\varTermd\varTerma
(\varTermt_1\conTit{\varTerms}{\varTermr})
(\varTermt_2\conTit{\varTerms}{\varTermr})}(\varnatN)\\
&+&\len{
\varTerms\varTermd\varTerma
(\varTermt_1\conTit{\varTerms}{\varTermr})
(\varTermt_2\conTit{\varTerms}{\varTermr})}
\\[1ex]
=&&\pollen{\varTerms}(\varnatN)+\pollen{\varTermd}(\varnatN)
+\pollen{\varTerma}(\varnatN)\\
&+&\pollen{\varTermt_1}(\varnatN)+\varnatn_1\cdot\pollen{\varTerms}(\varnatN)
                  +(\varnatn_1+1)\cdot\pollen{\varTermr}(\varnatN)
               +\varnatn_1\cdot\len{\varTerms}+\varnatn_1\cdot\len{\varTermr}\\
&+&\pollen{\varTermt_2}(\varnatN)+\varnatn_2\cdot\pollen{\varTerms}(\varnatN)
                  +(\varnatn_2+1)\cdot\pollen{\varTermr}(\varnatN)
               +\varnatn_2\cdot\len{\varTerms}+\varnatn_2\cdot\len{\varTermr}\\

&+&\len{\varTerms}+\len{\varTermd}+\len{\varTerma}+\len{\varTermt_1}
+\len{\varTermr}+\len{\varTermt_2}+\len{\varTermr}
\end{array}$$
which is strictly smaller.

It should be noted that, once iteration on labeled trees is available,
``divide and conquer'' recursion can be implemented: Starting from a
``divide function''
\[
\funF\colon\varTypeA\typeLinTo\typeL{\varTypeA}\typeLinTo\varTypeA
\typeTensor\typeL{\varTypeA}\typeTensor\typeL{\varTypeA}
\]
we can construct a function
$\tilde{\funF}\colon\typeL{\varTypeA}\typeLinTo\typeT{\varTypeA}{
\typeL{\varTypeA}}$ with
$$\begin{array}{lcl}
\tilde{\funF}
(\nil{\varTypeA}) &=& \emptyTree{\varTypeA}{\typeL{\varTypeA}}\nil{\varTypeA}
\\
\tilde{\funF} 
(\cons{\varTypeA}\varTermd^{\typeD}a^{\varTypeA}\varL) &=&
\consTree{\varTypeA}{\typeL{\varTypeA}}\varTermd a^{\prime}\varL_1\varL_2
\qquad\mbox{if~}\funF a\varL=a^{\prime}\typeTensor\varL_1\typeTensor\varL_2
\end{array}$$
To do so we first construct via iteration over lists a function 
$\funG\colon\typeL{\varTypeA}
\typeLinTo\typeL{\varTypeA}\typeCross\typeT{\varTypeA}{\typeL{\varTypeA}}$
with $\funG(\varL)=\conPair{\varL}{\tilde{\funF}\varL}$.
Iterating a function that applies $\tilde{\funF}$ to (the label of) every
leaf of a tree sufficiently (i.e., linearly)
often~\cite[\S4.3]{Hofmann99} we can define a function
$\funF^{D}\colon\typeL{\varTypeA}\typeLinTo\typeT{\varTypeA}{\typeL
{\varTypeA}}$ such that
$$\begin{array}{lcl}
\funF^D(\nil{\varTypeA}) &=& \emptyTree{\varTypeA}{\typeL{\varTypeA}}
(\nil{\varTypeA})\\
\funF^D(\cons\varTermd^{\typeD}a^{\varTypeA}\varL) &=&
\consTree{\varTypeA}{\typeL{\varTypeA}}\varTermd a^{\prime} (\funF^D\varL_1)
(\funF^D\varL_2)
\qquad\mbox{if~}\funF a\varL=a^{\prime}\typeTensor\varL_1\typeTensor\varL_2
\end{array}$$
A final iteration over the created tree with an appropriate ``conquer
function'' and an appropriate initial value finishes the
implementation of ``divide and conquer'' recursion in the present
system.

Following these lines, we can, for example, specify \texttt{quicksort}
by providing
\begin{itemize}
\item a divide-function of type
$\varTypeA\typeLinTo\typeL{\varTypeA}
\typeLinTo\varTypeA\typeTensor\typeL{\varTypeA}\typeTensor
\typeL{\varTypeA}$ splitting a list into two sublists; one with the
elements larger than the given one and one with smaller elements
\begin{align*}
\lambda\varx^{\varTypeA},\varl^{\typeL{\varTypeA}}\quad\varl&
\{\lambda\vary_0^{\typeD},\vary_1^{\varTypeA},\varp^{\varTypeA\typeTensor
\typeL{\varTypeA}\typeTensor{\varTypeA}}\quad\varp(\lambda\varx^{\varTypeA},
\varp^{\typeL{\varTypeA}\typeTensor\typeL{\varTypeA}}\quad\varp(\lambda\varl^
{\prime},\varl^{\prime\prime}\\
&\leqbsp{\varTypeA}\varx\vary_1
\begin{array}[t]{l}
(\lambda\varz^{\typeBool},\varp^{\varTypeA
\typeTensor\varTypeB}\qquad\varp(\lambda\varx^{\varTypeA},\vary_1^{\varTypeA}\\
z\langle\begin{array}[t]{l}\conTens{\varTypeA}{\typeL{\varTypeA}\typeTensor
\typeL{\varTypeA}}\varx(\conTens{\typeL{\varTypeA}}{\typeL{\varTypeA}}
(\cons{\varTypeA}\vary_0\vary_1\varl^{\prime})\varl^{\prime\prime}),\\
\conTens{\varTypeA}{\typeL{\varTypeA}\typeTensor\typeL{\varTypeA}}\varx
(\conTens{\typeL{\varTypeA}}{\typeL{\varTypeA}}\varl^{\prime}
(\cons{\varTypeA}\vary_0\vary_1\varl^{\prime\prime}))\rangle
))) \}
\end{array} \end{array} \\
&\conTens{\varTypeA}{\typeL{\varTypeA}\typeTensor\typeL{\varTypeA}}\varx
(\conTens{\typeL{\varTypeA}}{\typeL{\varTypeA}}\nil{\varTypeA}\nil{\varTypeA}
)
\end{align*}
\item a conquer function of type
$\typeD\typeLinTo\varTypeA\typeLinTo
\typeL{\varTypeA}\typeLinTo\typeL{\varTypeA}\typeLinTo\typeL{\varTypeA}$
taking two sorted lists and joining them in the correct way with the
middle element
$$
\lambda\varx^{\typeD},\vary^{\varTypeA},\varl^{\prime}{}^{\typeL{\varTypeA}},
\varl^{\prime\prime}{}^{\typeL{\varTypeA}}\quad
\mathtt{app}\varl^{\prime}(\cons{\varTypeA}\varx\vary\varl^{\prime\prime})
$$
where $\mathtt{app}$ is the append function, defined as usual
\item an initial case for the empty list, 
$\nil{\varTypeA}$
\end{itemize}

However, as in this \emph{implementation} of ``divide and conquer''
recursion for each unfolding step the whole (intermediate) tree
structure has to be traversed, it is less efficient than a ``native''
variant of ``divide and conquer'', but, as shown, still in polynomial
time.
\end{subsection}
\end{section}



\begin{thebibliography}{10}
\bibitem{AehligSchwichtenberg00}
Klaus Aehlig and Helmut Schwichtenberg.
\newblock A syntactical analysis of non-size-increa\-sing polynomial time
  computation.
\newblock In {\em Proceedings 15'th Symposium on Logic in Computer Science
  (LICS 2000)}, pages 84--91, 2000.

\bibitem{BeckmannWeiermann00}
Arnold Beckmann and Andreas Weiermann.
\newblock {Characterizing the elementary recursive functions by a fragment of
  G\"odel's $T$}.
\newblock {\em Archive for Ma\-the\-ma\-ti\-cal Logic}, 39(7):475--492, 2000.

\bibitem{BellantoniCook92}
Stephen Bellantoni and Stephen Cook.
\newblock A new recursion-theoretic characterization of the polytime functions.
\newblock {\em Computational Complexity}, 2:97--110, 1992.

\bibitem{BellantoniNigglSchwichtenberg00}
Stephen Bellantoni, Karl-Heinz Niggl, and Helmut Schwichtenberg.
\newblock Higher type recursion, ramification and polynomial time.
\newblock {\em Annals of Pure and Applied Logic}, 104:17--30, 2000.

\bibitem{Caseiro97}
Vuokko-Helena Caseiro.
\newblock {\em Equations for Defining Poly-Time Functions}.
\newblock PhD thesis, Department of Computer Science, University of Oslo, 1997.
\newblock Available at www.ifi.uio.no/~ftp/publications/research-reports/.

\bibitem{Cook71}
Stephen~A. Cook.
\newblock Characterizations of push-down machines in terms of time bounded
  computers.
\newblock {\em Journal of the Association for Computing Machinery}, 18:4--18,
  1971.

\bibitem{Goedel58}
Kurt G{\"o}del.
\newblock {{\"U}ber eine bisher noch nicht ben{\"u}tzte Erweiterung des finiten
  Standpunkts}.
\newblock {\em Dialectica}, 12:280--287, 1958.

\bibitem{Hofmann01}
Martin Hofmann.
\newblock A type system for bounded space and functional in-place update.
\newblock To appear: Nordic Journal of Programming. An extended abstract has
  appeared in `Programming Languages and Systems' (Proc. ESOP 2000), G. Smolka,
  ed., Springer LNCS, 2000.

\bibitem{Hofmann98b}
Martin Hofmann.
\newblock Typed lambda calculi for polynomial-time computation.
\newblock {Habilitation thesis, Mathematisches Institut, TU Darmstadt, Germany.
  Available under www.dcs.ed.ac.uk/home/mxh/habil.ps.gz}, 1998.

\bibitem{Hofmann99}
Martin Hofmann.
\newblock Linear types and non-size-increasing polynomial time computation.
\newblock In {\em Proceedings 14'th Symposium on Logic in Computer Science
  (LICS'99)}, pages 464--473, 1999.

\bibitem{Hofmann00}
Martin Hofmann.
\newblock Linear types and non-size-increasing polynomial time computation.
\newblock To appear: Theoretical Computer Science, 2000.

\bibitem{JoachimskiMatthes99}
Felix Joachimski and Ralph Matthes.
\newblock {Short proofs of normalisation for the simply-typed
  $\lambda$-calculus, permutative conversions and G{\"o}del's $T$}.
\newblock To appear: Archive for Mathematical Logic, 1999.

\bibitem{Jones97}
Neil~D. Jones.
\newblock {\em Computability and Complexity from a Programming Perspective}.
\newblock MIT Press, 1997.

\bibitem{Jones99}
Neil~D. Jones.
\newblock {LOGSPACE and PTIME characterized by programming languages}.
\newblock {\em Theoretical Computer Science}, 228:151--174, 1999.

\bibitem{Jones01}
Neil~D. Jones.
\newblock {The expressive power of higher-order types or, life without CONS}.
\newblock {\em J. Functional Programming}, 11(1):55--94, 2001.

\bibitem{Leivant91}
Daniel Leivant.
\newblock A foundational delineation of computational feasibility.
\newblock In {\em Proceedings 6'th Symposium on Logic in Computer Science
  (LICS'91)}, 1991.

\bibitem{Leivant99}
Daniel Leivant.
\newblock Applicative control and computational complexity.
\newblock In J.~Flum and M.~Rodriguez-Artalejo, editors, {\em Computer Science
  Logic (Proceedings of the Thirteenth CSL Conference)}, number 1683 in LNCS,
  pages 82--95. Springer Verlag, Berlin, Heidelberg, New York, 1999.

\bibitem{LeivantMarion93a}
Daniel Leivant and Jean-Yves Marion.
\newblock Lambda calculus characterization of poly-time.
\newblock {\em Fundamenta Informaticae}, 19:167--184, 1993.

\bibitem{Simmons88}
Harold Simmons.
\newblock The realm of primitive recursion.
\newblock {\em Archive for Ma\-the\-ma\-ti\-cal Logic}, 27:177--188, 1988.

\end{thebibliography}
\end{document}